\documentclass{aa}
\usepackage{natbib} \usepackage{graphicx}
\bibpunct{(}{)}{;}{a}{}{,}
\begin{document}

\title{Relationships between magnetic foot points and G-band bright structures}

\author{R.Ishikawa\inst{1,2} \and S.Tsuneta\inst{2} \and Y.Kitakoshi\inst{1,2} 
            \and Y.Katsukawa\inst{2} \and J.A.Bonet\inst{3} \and S.Vargas Dom\'inguez\inst{3}
                  \and L.H.M.Rouppe van der Voort\inst{4,5} \and Y.Sakamoto\inst{1,2} \and T.Ebisuzaki\inst{6}}
\institute{Department of Astronomy, University of Tokyo, Hongo, Bunkyo-ku, Tokyo 113-0033, Japan
 \and National Astronomical Observatory of Japan, 2-21-1 Osawa, Mitaka, Tokyo 181-8588, Japan
   \and Instituto de Astrof\'isica de Canarias, 38205 La Laguna, Tenerife, Spain
      \and Institute of Theoretical Astrophysics, University of Oslo, P.O. Box 1029 Blindern, N-0315 Oslo, Norway
        \and Center of Mathematics for Applications, University of Oslo, P.O. Box 1053 Blindern, N--0316 Oslo, Norway
          \and Astrophysical Computing Center, RIKEN, 2-1 Hirosawa, Wako, Saitama 351-0198, Japan.}
\date{Received/Accepted 24 April 2007}

\abstract
{}
{Magnetic elements are thought to be described by flux tube models, and are well reproduced by MHD simulations. 
However, these simulations are only partially constrained by observations. 
We observationally investigate the relationship between G-band bright points and magnetic structures to clarify conditions, which make magnetic structures bright in G-band.}
{The G-band filtergrams together with magnetograms and dopplergrams were taken for a plage region covered by abnormal granules as well as ubiquitous G-band bright points,  using the Swedish 1-m Solar Telescope (SST) under very good seeing conditions.}
{High magnetic flux density regions are not necessarily associated with G-band bright points.
We refer to the observed extended areas with high magnetic flux density as magnetic islands to separate them from magnetic elements.
We discover that G-band bright points tend to be located near the boundary of such magnetic islands.
The concentration of G-band bright points decreases with inward distance from the boundary of the magnetic islands.
Moreover, G-band bright points are preferentially located where
magnetic flux density is higher, given the same distance from the boundary.
There are some bright points located far inside the magnetic islands.
Such bright points have higher minimum 
magnetic flux density at the larger inward distance from the boundary.
Convective velocity is apparently reduced for such 
high magnetic flux density regions regardless of whether they are populated by G-band bright points or not.
The magnetic islands are surrounded by downflows.}
{These results suggest that high magnetic flux density, as well as efficient heat transport from the sides or beneath, are required to make magnetic elements bright in G-band.}
\maketitle
\section{Introduction}

\begin{figure*} 
 \centering
  \resizebox{18cm}{!}{\includegraphics[angle=90]{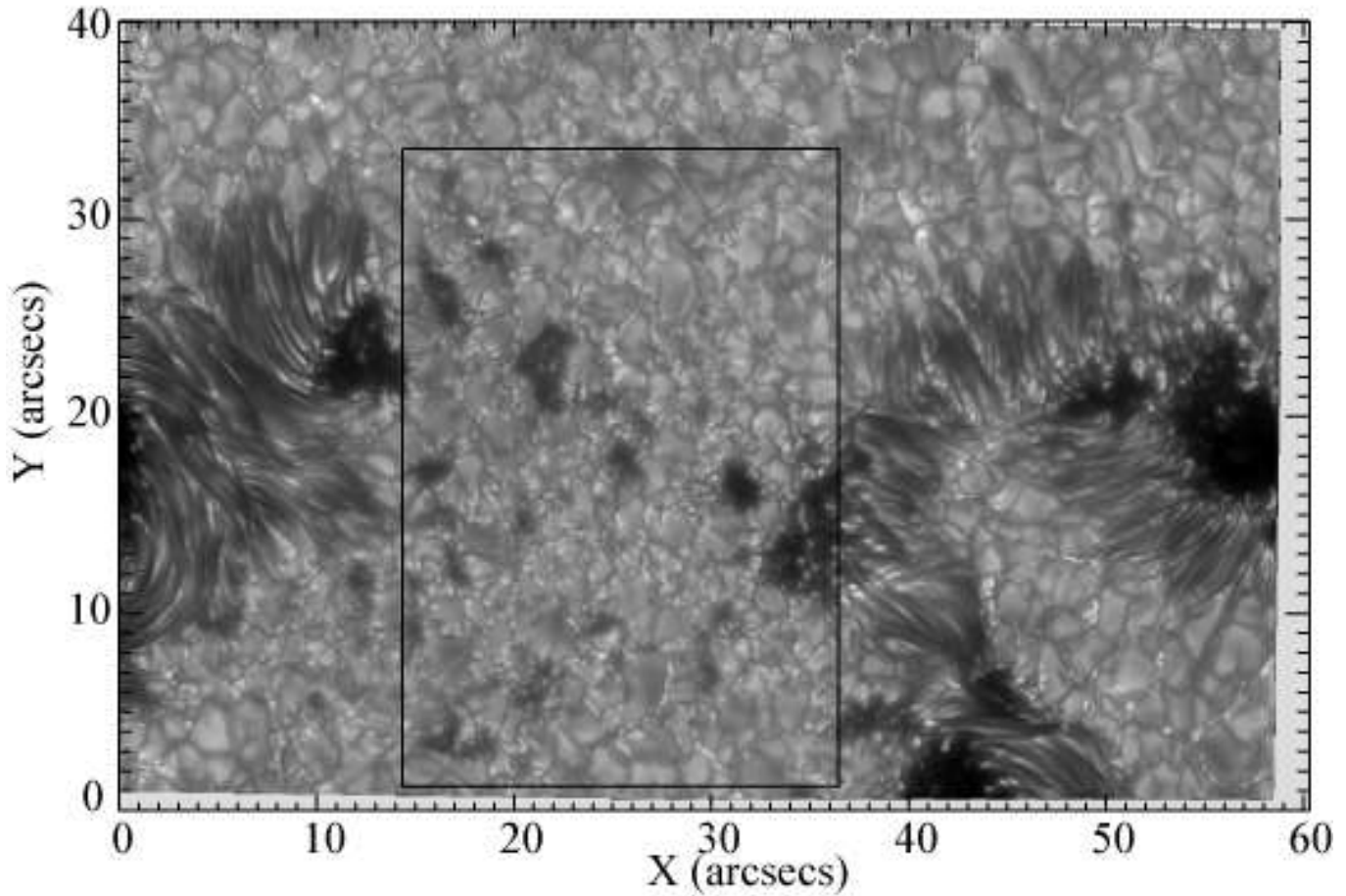}}
  \caption{G-band filtergram taken at 08:45:58 UT on 9 July 2005. The inner box indicates the region analyzed in this paper.}
  \label{whole_img}
\end{figure*}

\begin{figure*} 
 \centering
  \includegraphics[width=10cm, angle=90]{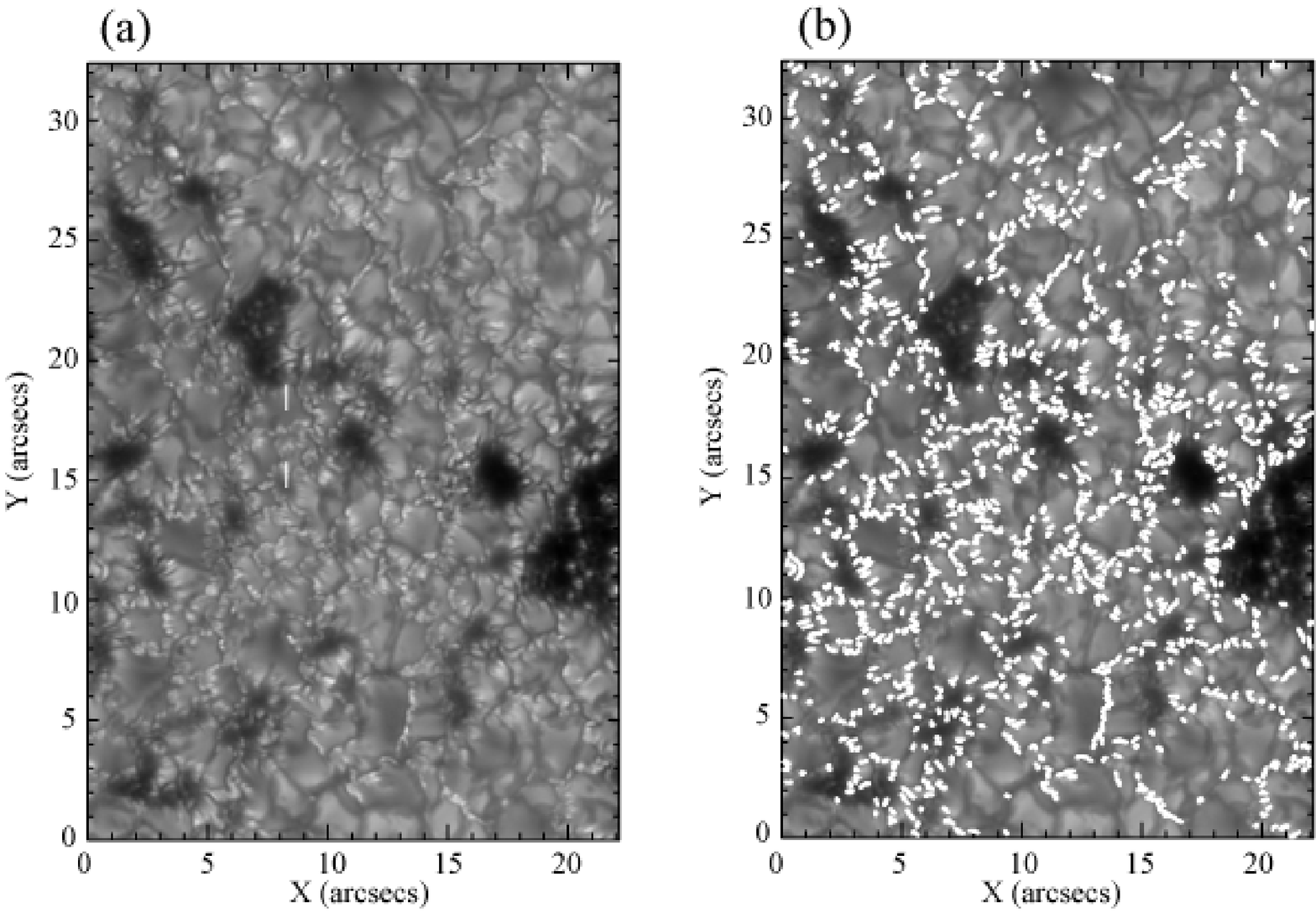}
  \caption{$a):$ The region marked by the \textbf{box} in fig.\ref{whole_img}. The white lines denote a cut along which the intensity profile is shown in fig.\ref{ex_plot}. $b):$ G-band bright points are selected and marked in white. Bright dots inside pores are not identified as G-band bright points.}
  \label{plage_gbp}
\end{figure*}

Magnetic fields on the Sun have various spatial and time scales.
Sunspots have the largest scale size ($\sim10^{4-5}\mathrm{km}$) with long lifetimes ($\sim1\mathrm{month}$) and very high contrast, 
while magnetic elements often observed in the Fraunhofer G-band (the CN band, the wing of Ca H and K, and continuum) are the smallest resolved structures, and are believed to be building blocks of active regions and the quiet Sun magnetic network.
In the 1970s, filigree structures in the intergranular lanes and faculae observed at the solar limb were reported \citep{dunn1973, mehl1974}, and these small bright points were believed to be a manifestation of elemental magnetic fields \citep{spruit1976, spruit1977}.
\citet{beckers1968} originally observed small magnetic features called magnetic knots with magnetic field strength between 1000-2000 G.
Recent high-resolution observations have been able to resolve these magnetic knots into smaller magnetic elements.
The question of how such small structures with strong magnetic flux density and with short lifetime exist is one of the central issues in solar
magnetohydrodynamics.
Direct magnetic observations of the magnetic elements are, however, difficult because of their small spatial scales and short evolution time-scales.
One of the indirect methods for tracing the magnetic elements is to use their association with G-band bright points.

We see deeper layers of the Sun in the magnetic elements than in the surrounding non-magnetic photosphere, because the lower density due to magnetic pressure decreases the optical depth.
Thus, we see hotter and brighter layers inside the magnetic elements that are thin enough to be heated by the surrounding environment \citep{spruit1976, spruit1977}.
The basic concept has been further elaborated by many authors in recent years \citep{sanchezal2001, steiner2001, carlsson2004, keller2004,  uitenbroektritschler2006, leenaartsetal2006}.  
Recent high resolution observations have revealed various properties of the G-band bright points such as the size \citep{berger1995, bovelet2003, wieh2004}, the apparent shape \citep{berger1995, bovelet2003}, the mean velocity \citep{berger1998, bovelet2003, nisenson2003}, and the mean lifetime \citep{berger1998, nisenson2003}.
G-band bright points in the form of a thin sheet, which appears not to be resolved into smaller G-band bright points, have been seen \citep{berger2004, rouppe2005}.

These studies are carried out, assuming that the G-band bright points directly correspond to magnetic elements. 
\citet{keller1992n} reported that magnetic flux concentrations with diameters of $\approx 200\mathrm{km}$ were identified with the bright points using speckle interferometry at the Swedish 50-$\mbox{cm}$ telescope.
The G-band bright points in the intergranular lane are cospatial and comorphous with magnetic elements \citep{berger2001}.
However, few studies so far statistically show the relationship between G-band bright points and magnetic concentration:
it may be true that any G-band bright points have strong magnetic fields, but vice versa is apparently not the case.
The difficulty of studying such correspondence stems from the relatively low spatial resolution of the spectropolarimetric measurements and magnetograms, compared to the wide band imaging.

Fig.\ref{whole_img} shows the region consisting of small pores and many abnormal granules. 
G-band bright points in the region appear to be smaller and have higher concentration than those around the sunspot located in the right corner of the field of view. 
The simultaneous magnetogram shows multiple magnetic regions. 
In this paper, we investigate the relationship between such extended magnetic regions and G-band bright points, and observationally address the question of why the bright points appear bright.

\section{Observations and data analysis}
\subsection{Observations}
The observations were carried out during an international observing campaign involving the Swedish 1-m Solar Telescope (SST), the Dutch Open Telescope, the German Vacuum Tower Telescope, and the TRACE satellite during 2 weeks in July 2005.
NOAA AR 10786 located at heliographic coordinates N14, W25 was observed with the SST \citep{scharmer2003SST} from 7:47 to 9:10 UT on 9 July 2005 under good and stable seeing conditions. 
We analyze magnetograms and Dopplergrams taken in the \ion{Fe}{i}~630.25~nm line with the Solar Optical Universal Polarimeter \citep[SOUP,][]{title1981},
 and G-band images obtained with an interference filter centered on 430.56~nm with the bandpass of 1.1~nm. 

The data  quality is extremely high due to the SST adaptive optics system \citep[AO,][]{scharmer2003AO} and post-processing with the Multi-Object Multi-Frame Blind Deconvolution technique \citep[MOMFBD,][]{noort2005}.
The G-band images and nearby continuum phase diversity image-pairs were simultaneously recorded in three channels at a rate of 1.8 frames~s$^{-1}$.
Manual frame selection was applied immediately after the observations, using the image contrast as a measure for selecting images for final storage.
Thus, only those images were selected, for which the AO system was actively compensating for seeing degradation.

Typically, 18 images per channel (i.e. 3 $\times$ 18) were combined by the MOMFBD algorithm to render a pair of simultaneous G-band
and nearby continuum restored images. 
Following this procedure, two time series of restored images for G-band and nearby continuum with a cadence of 10~s were obtained.
The field of view (FOV) for the G-band and nearby continuum images is $60\arcsec$ $\times$ $40\arcsec$ with a pixel size of 0\farcs041.
From an analysis of the power spectrum we conclude that the achieved spatial resolution is very close to the diffraction limit: 0\farcs1.

The SOUP filter was running a program switching between the blue and red wing of the \ion{Fe}{i}~630.25~nm line at $\pm 5$~pm offset, and 10 left- and 10 right-hand circular polarization (LCP and RCP) images were recorded for each line position,
alternating polarization states during the read-out time of the camera (0.82~s).
The total time needed to acquire all images of the four types was 33~s. 
The MOMFBD restoration was performed from subsets of 40 SOUP images (all states and line positions) and 40 simultaneously recorded wide-band images.
The wide-band images were recorded by splitting off 10\% of the light after the SOUP pre-filter (0.8~nm passband centered on 630.25~nm), and are used as an anchor channel.
Since the MOMFBD restoration uses a camera alignment calibration involving pinhole images,
the four restored SOUP images are very accurately aligned to the wide-band anchor channel.
Magnetograms and Dopplergrams can therefore be constructed without any significant level of seeing cross-talk.
See \citet{noort2005} for more details.
The FOV of the restored SOUP images is 90$\arcsec$ $\times$ 60$\arcsec$ (plate scale 0\farcs061~pixel $^{-1}$). 
We estimate that the spatial resolution in the magnetogram is about 0\farcs2.
The exposure time for the individual SOUP images was 40~ms. For the magnetogram based on the MOMFBD restored images this yields to an effective exposure time of 1.6~s.

\begin{figure} 
 \centering
  \includegraphics[width=4.5cm, angle=90]{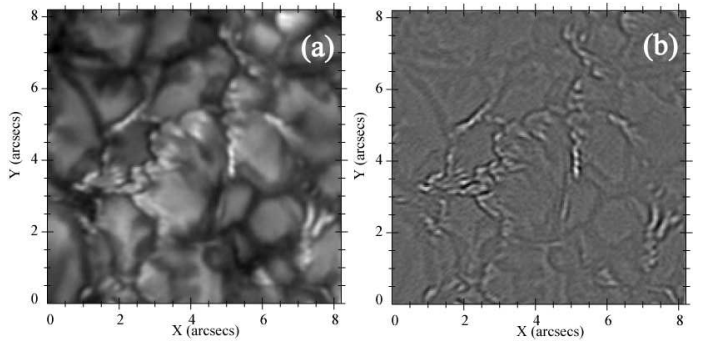}
  \caption{$a):$ The subregion of the plage region shown in fig.\ref{plage_gbp}. $b):$ The Laplacian-filtered image.}
  \label{laplacian_filt}
\end{figure}
\begin{figure*}
 \centering
  \includegraphics[width=10cm, angle=90]{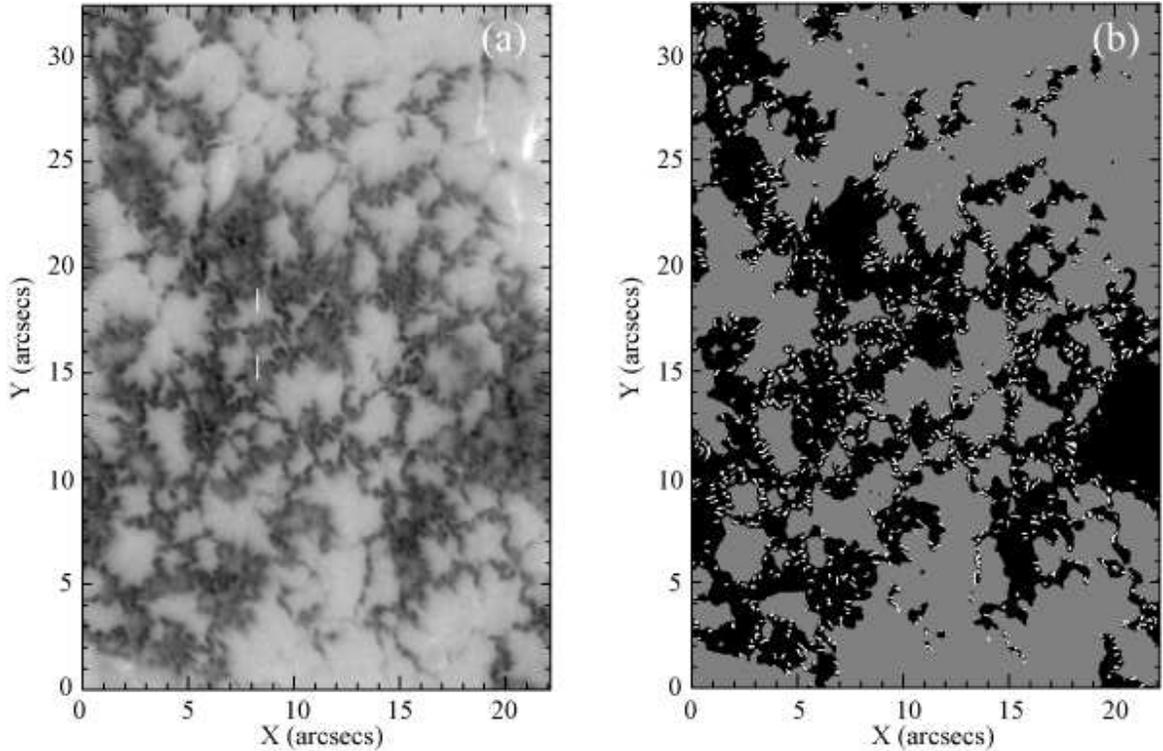} 
  \caption{$a)$ Magnetogram corresponding to the region marked by the rectangle in fig.\ref{whole_img} (08:45:58 UT). The gray scale ranges from -2795 (black) to 1000 (white) Gauss. The white lines denote a cut along which an intensity profile is shown in fig.\ref{ex_plot}. $b)$ Magnetic islands are marked in black and G-band bright points in magnetic islands are marked in white.}
  \label{plage_mag}
\end{figure*} 
\subsection{Data reduction}
Magnetograms, $M$, and Dopplergrams, $V$, were computed as follows:

\begin{equation}
M = \frac{1}{2}\left[\frac{I^{blue}_{RCP}-I^{blue}_{LCP}}{I^{blue}_{RCP}+I^{blue}_{LCP}}+\frac{I^{red}_{LCP}-I^{red}_{RCP}}{I^{red}_{RCP}+I^{red}_{LCP}}\right],\label{M}
\end{equation}

\begin{equation}
V = \frac{1}{2}\left[\frac{I^{blue}_{RCP}-I^{red}_{RCP}}{I^{blue}_{RCP}+I^{red}_{RCP}}+\frac{I^{blue}_{LCP}-I^{red}_{LCP}}{I^{blue}_{LCP}+I^{red}_{LCP}}\right]\label{M},\label{V}\end{equation}
and the Doppler velocity $v$ is given by

\begin{equation}
v\left[\mathrm{km\ s^{-1}}\right] = f_{c}V,\label{v}
\end{equation}
where $f_{c}$ is a constant to convert Doppler signal to the Doppler velocity.
The line-of-sight magnetic flux density $B$ is proportional to the magnetogram signal $M$:

\begin{equation}
B\left[\mathrm{Gauss}\right] = \alpha M.\label{B}
\end{equation}

Magnetic flux density derived here actually corresponds to magnetic flux of a pixel.  
It is equivalent to the magnetic flux density (or magnetic field strength) only when the pixel is occupied by magnetic atmosphere with a filling factor of 1.
We adopt 16551 $\mathrm{Gauss}$ for $\alpha$ \citep{berger2004}.
From a relatively quiet, noise dominated, region in the FOV, we estimate that the sensitivity is 0.3\%, or 53~Gauss.
The magnetographic interpretations are subject to the effects of such things as different thermal conditions, gradients of plasma flows along the line-of-sight, and incomplete angular resolution. 
Equation \ref{B} may not be valid above approximately 1 kG due to the Zeeman saturation for the wavelength offset of 5~pm. 
Note, however, that the magnetogram signals would still differentiate the relative magnitude of the magnetic flux density beyond the linear regime. 
This is not the case for some of the pores which have strong magnetic flux density. Indeed, some of the pores should have had larger magnetic flux density in figure \ref{plage_mag}.

The constant $f_{c}$ is estimated to be $\sim$ 10$\mathrm{\ km\ s^{-1}}$ by comparing the Doppler signal with the mean line-of-sight Doppler velocity of granules ($\sim1\mathrm{km\ s^{-1}}$).
The plate scale for the wide-band images and the SOUP images are adjusted to that of the G-band images,
and the G-band images are aligned with sub-pixel accuracy to the wide-band images.
We select the MOMFBD-restored G-band image with the highest contrast,
and the Dopplergram and the magnetogram nearest to the G-band image.
We here concentrate on the region with size of $22\arcsec \times 32\arcsec$ to analyze G-band bright points in the plage region (fig.\ref{plage_gbp} $a$).

\begin{figure}
 \centering
  \includegraphics[width=6.5cm, angle=90]{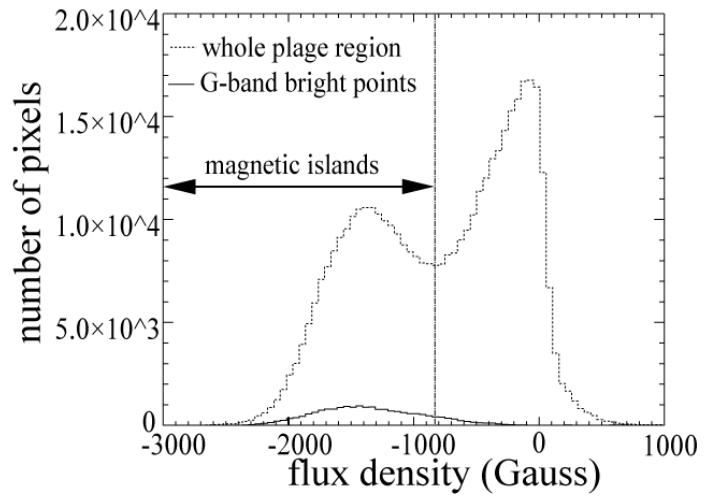} 
  \caption{Histogram of magnetic flux density for the whole region including G-band bright points in fig.\ref{plage_gbp} $a$ and fig.\ref{plage_mag} $a$ (dotted line) and that for the G-band bright points identified in fig.\ref{plage_gbp} $b$ (solid line). The arrow indicates the range for magnetic islands.}
  \label{maghist}
\end{figure} 

\subsection {G-band bright points and Magnetic Islands}
The G-band bright points in the region are not always round, but are elongated or have the form of a thin sheet (fig.\ref{plage_gbp}). 
Though there are many such structures in the region, we refer to those bright structures in the G-band as the G-band bright points in this paper.

A simple threshold in intensity cannot properly identify the G-band bright points,
some of which have low contrast with respect to the surrounding granular patterns.
Thus, we employ a Laplacian filter \citep{kano1995lp}, which provides a second derivative of an image.
The Laplacian-filter procedure is given by

\begin{eqnarray}
\nabla^{2}I(x,y)&=&-[\{I(x+1,y)-I(x,y)\} \nonumber\\
&\;&\;\;-\{I(x,y)-I(x-1,y)\} \nonumber\\
&\;&\;\;+\{I(x,y+1)-I(x,y)\} \nonumber\\
&\;&\;\;-\{I(x,y)-I(x,y-1)\}] \nonumber \\
&=&-\{I(x+1,y)+I(x-1,y) \nonumber\\
&\;&\;\;+I(x,y+1)+I(x,y-1)-4I(x,y)\} .\label{laplacian}
\end{eqnarray}

The filter can highlight local intensity peaks and valleys, and allows us to detect any structure even with small enhancement.
Figure \ref{laplacian_filt} shows a sub-region of the Laplacian-filtered G-band image.
Small bright features have very high contrast, while edges of bright granules do not have high contrast in the Laplacian-filtered G-band image. 
We adjust a threshold for the Laplacian-filtered images to correctly pick up enhanced structures with respect to the surroundings, and
identify the regions with the area larger than 9~pixels as G-band bright points. 
Figure \ref{plage_gbp} shows that the Laplacian filter almost perfectly picks up the G-band bright points.
Note that the G-band bright points inside pores are excluded.

The magnetogram (fig.\ref{plage_mag} $a$) shows the extended regions with strong magnetic flux density. 
Figure \ref{maghist} is the histogram of magnetic flux density for the entire plage region (including G-band bright points) and for only the G-band bright points. 
There are intense magnetic fields not associated with the G-band bright points. 
The pixels with strong magnetic flux density form an extended area not necessarily in a form of elemental magnetic fields.
Thus, we call these extended magnetic areas $magnetic$ $islands$.
Magnetic islands refer to any extended magnetic region that may or may not contain G-band bright points, pores, and any other magnetic features such as ribbons and flowers.
Magnetic islands are larger in size than ribbons, flowers, and micro-pores discussed in \citet{berger2004} and \citet{rouppe2005}. 
Those are typically of the size of granules.
A magnetic island here is defined to be an area whose absolute flux density is higher than the threshold value at which the number of pixels is minimal in figure \ref{maghist}.
The actual value of the threshold thus determined is not critical.
We adopt the threshold, simply because this clearly defines the apparent boundary of the magnetic islands. 
Magnetic islands include G-band bright points, pores, and an umbra on the right side of figure \ref{plage_gbp}. 
There are multiple magnetic islands without pores.

Figure \ref{plage_mag} $b$ shows the location of the G-band bright points (marked by white lines) in the magnetic islands (black area).
The figure clearly shows that G-band bright points are located near the boundary of the magnetic islands.

\section{Results}
\subsection{Spatial distribution of G-band bright points in plage region}
We confirm that all the G-band bright points have magnetic fields as shown in figure \ref{maghist}.
It is also clear from the figure that all the magnetic fields are not necessarily associated with G-band bright points. 
This simple figure raises the question
as to what the necessary condition for G-band bright points is.
In section, 2.3, we conclude from figure \ref{plage_mag} $b$ that G-band bright points tend to be located near the boundaries of the magnetic islands.
We quantitatively examine where  in the magnetic islands G-band bright points are located.

\begin{figure}
 \centering
  \includegraphics[width=3.5cm, angle=90]{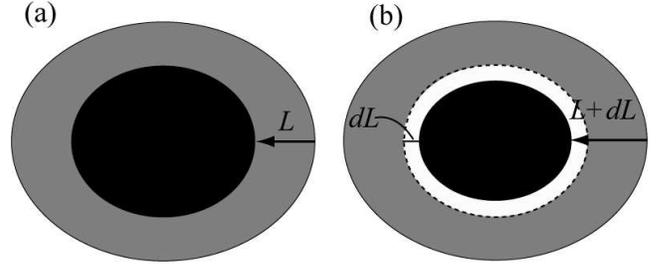}
  \caption{The schematic diagram of a magnetic island. $L$ and $L+dL$ are the inward distance from the boundary of the magnetic island as shown in the arrow.
See text for details.}
  \label{schematic_diagram}
\end{figure}

\begin{figure*}
 \centering
  \includegraphics[width=9cm, angle=90]{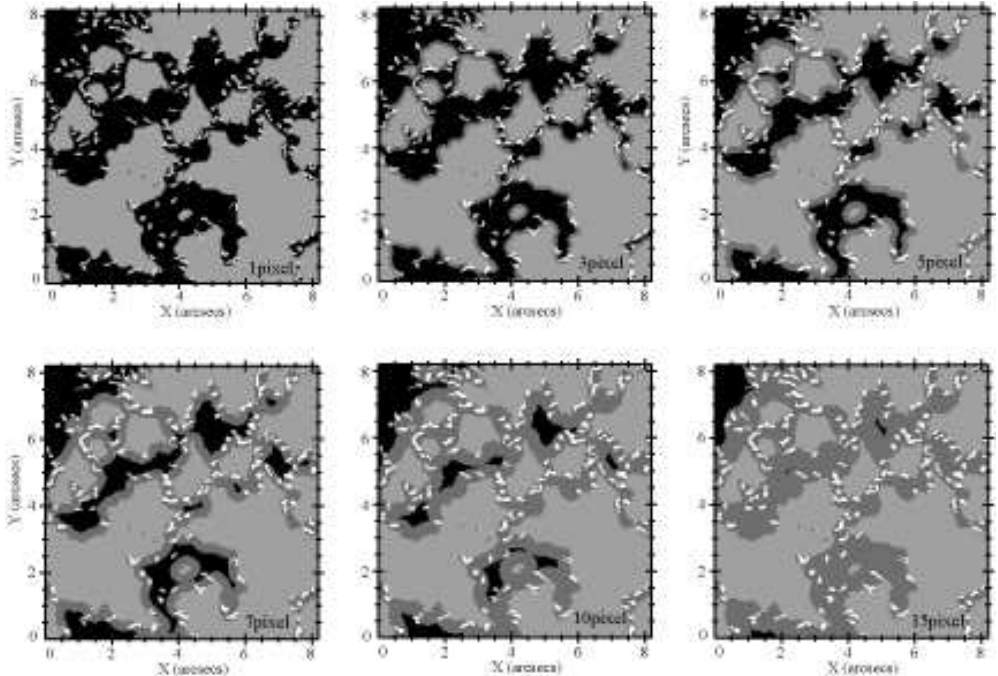}
  \caption{Eroded (from top left to bottom right) magnetic islands marked in black. Removed magnetic islands marked in dark gray. The number in lower right corner shows inward distance of erosion.}
  \label{erosion_example}
\end{figure*}

The area density $D(L)$ of the G-band bright points at an inward distance $L$ from the boundary of a magnetic island with width $dL$ (figure \ref{schematic_diagram}) is given by

 \begin{equation}
  D(L)=\frac{N_{gbp}(L)}{N_{all}(L)},\label{D}
 \end{equation}
where $N_{all}(L)$  are the number of pixels which make up the magnetic islands and $N_{gbp}(L)$ are the G-band bright points located in between $L$ and $L+dL$ at the inward distance $L$.

The procedure for obtaining $N_{i}(L)$ with $i=all,gbp$ is the following.
First, we obtain the total area of the black region (fig.\ref{schematic_diagram} $a$), $S_{all}(L)$, and the total area of the G-band bright points in the black region, $S_{gbp}(L)$.
Second, we obtain the total area of the smaller black region (fig.\ref{schematic_diagram} $b$), $S_{all}(L+dL)$, and the total area of the G-band bright points inside the smaller black region, $S_{gbp}(L+dL)$, by eroding with distance $dL$ ($\sim$1~pixel).
Note that we employ a mask to exclude the bright structures in the pores and the umbra.
Finally, the areas  $N_{i}(L)$ for $i=all,gbp$, are given by
\begin{equation}
  N_{i}(L)=S_{i}(L)-S_{i}(L+dL),\label{N}
\end{equation} 
and
\begin{equation}
 S_{i}(L)=\sum_{l \geq L} N_{i}(l),\label{S}
\end{equation}
The process of eroding magnetic islands is shown in figure \ref{erosion_example}.

\begin{figure} 
\includegraphics[width=6cm, angle=90]{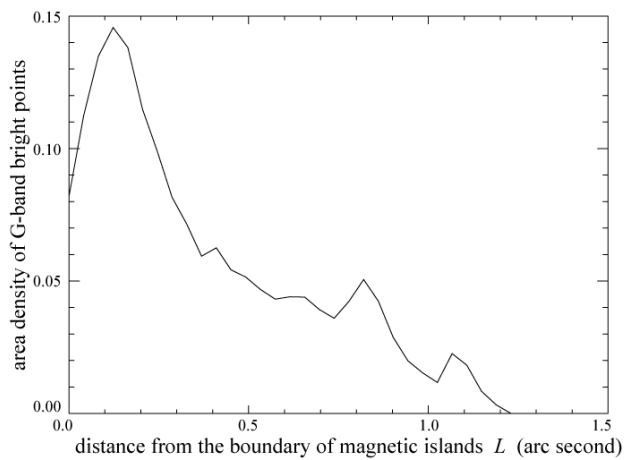}
  \caption{Area density $D(L)$ of the G-band bright points in magnetic islands as a function of inward distance $L$ from the boundary of magnetic islands.}
  \label{distance_gbpdensity}
\end{figure}

Figure \ref{distance_gbpdensity} shows the area density $D(L)$ of the G-band bright points derived in this way.
There are more G-band bright points near the boundary of the magnetic islands with a rapid decrease over $\sim0\farcs1$.
The inward distance of $0\farcs12$ ($3$ pixels) at which the area density of G-band bright points becomes maximum may depend slightly on the threshold used to define the magnetic islands. 

\begin{figure}
\includegraphics[width=6cm, angle=90]{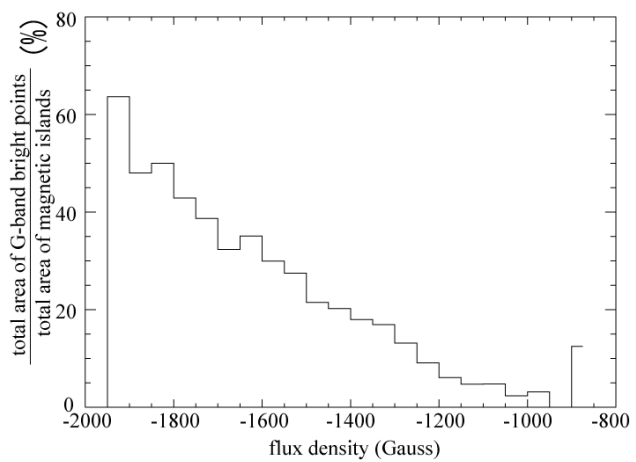}
  \caption{The histogram of G-band bright point density as a function of magnetic flux density at a distance of $0\farcs12$ (3 pixels) inward from the boundary of magnetic islands.}
  \label{flux_3th}
\end{figure}

Figure \ref{flux_3th} shows the area density of the G-band bright points for an inward radius $L$ of $0\farcs12$ with width $dL$ of $0\farcs041$ as a function of magnetic flux density. 
Regions with higher magnetic flux density are apparently in association with the G-band bright points at the same distance from the boundary.

\begin{figure} 
\resizebox{8cm}{!}{\includegraphics{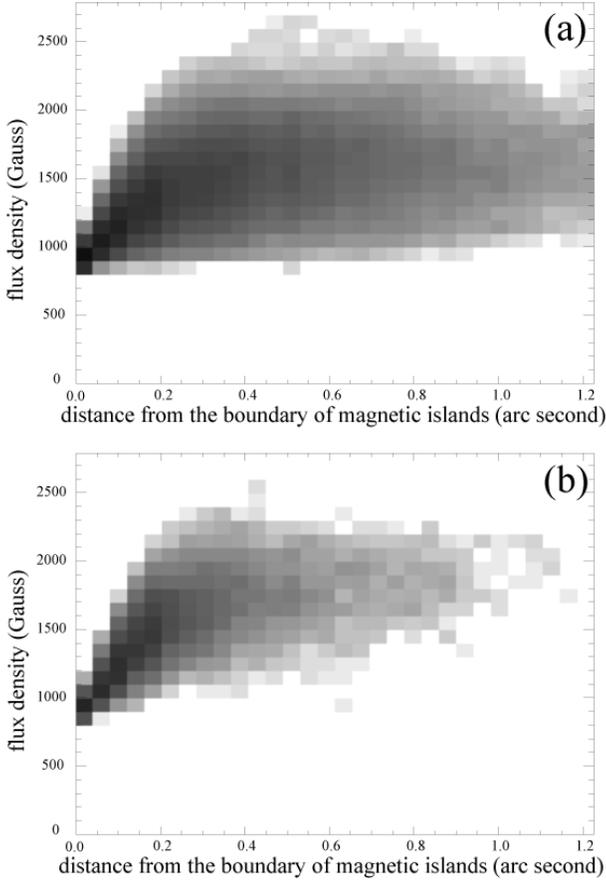}}
  \caption{A scatter plot of the magnetic flux density versus the inward distance from the boundary of magnetic islands for the magnetic islands ($a$) and for the G-band bright points ($b$).
The gray scale indicates the logarithmic number of data points.
Note that the magnetic flux density as derived from equation \ref{B} does not provide correct values for very high flux density due to Zeeman saturation for 50mA offset.}
  \label{distance_flux}
\end{figure}  

Figure \ref{distance_flux} $a$ and $b$ represent the magnetic flux density of the magnetic islands and of the G-band bright points as a function of the distance from the boundary of the magnetic islands respectively.
The magnetic flux density of the magnetic islands is almost uniformly distributed, and does not depend on the distance.  
However, the minimum flux density of G-band bright points does depend on the inward distance.
The G-band bright points away from the boundary have higher minimum magnetic flux density.

\begin{figure} 
\resizebox{8cm}{!}{\includegraphics[angle=90]{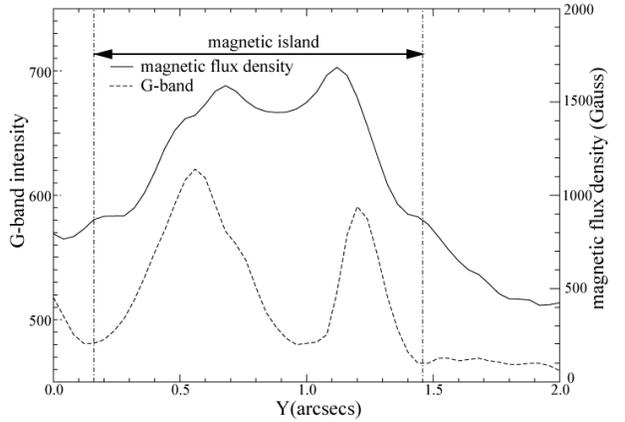}}
  \caption{Intensity plot along the white lines shown in fig.\ref{plage_gbp} $a$ and fig.\ref{plage_mag} $a$. The arrow indicates the region identified as a magnetic island.}
  \label{ex_plot}
\end{figure} 
 
Figure \ref{ex_plot} shows a cut across the magnetic island along the lines shown in figure \ref{plage_mag} $a$. 
This is similar to figures 7 and 9 in \citet{berger2004}.
Figure \ref{ex_plot} is, however, an example to show our statistical result that enhanced structures in the G-band are located near the edge of the magnetic island.
G-band bright points seem to prefer the neighborhood of the boundary of magnetic islands as well as higher magnetic flux density.

\subsection{Convection and Magnetic Islands}
\begin{figure} 
\resizebox{9cm}{!}{\includegraphics{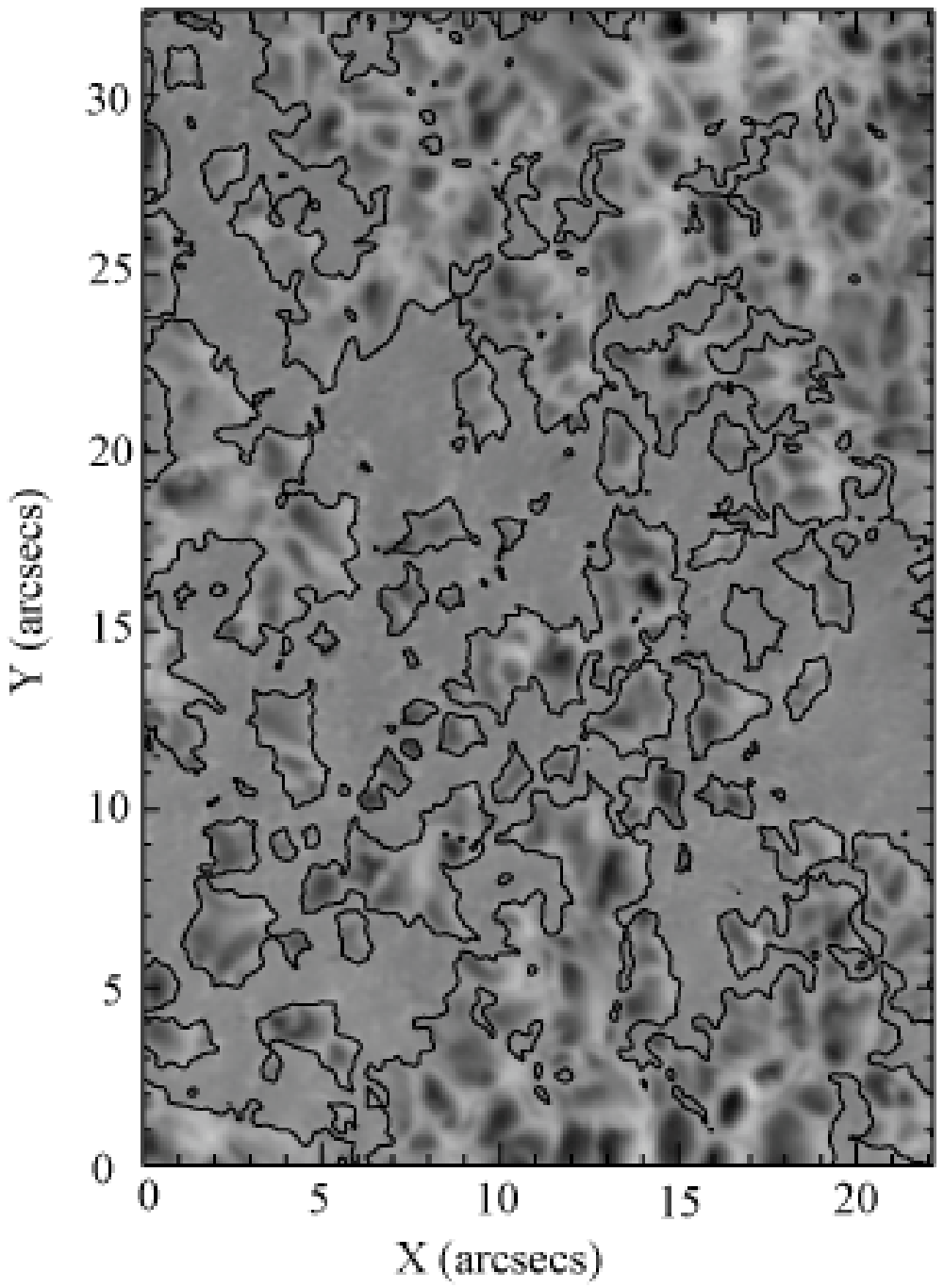}}
  \caption{Dopplergram corresponding to the region marked by the rectangle in fig\ref{whole_img} (08:45:58 UT). The black contour shows the boundary of magnetic islands (fig.\ref{plage_mag} $b$). The gray scale ranges from -2.3 km s$^{-1}$ (blue-shifted, black) to 2.3 km s$^{-1}$ (red-shifted, white).
The regions inside the magnetic islands have almost zero velocity.}
  \label{plage_dop}
\end{figure}  
\begin{figure} 
\resizebox{9cm}{!}{\includegraphics{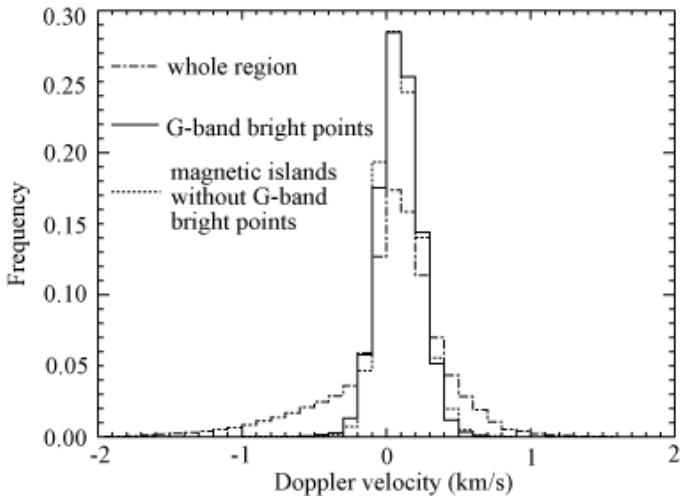}}
  \caption{Histogram of Doppler velocity for the whole region of fig.\ref{plage_gbp} $a$ (dash-dotted line), the G-band bright points (solid line) and magnetic islands without the G-band bright points (dotted line). Negative values are upflows (blue-shifted), positive downflows (red-shifted).}
  \label{dophist1}
\end{figure}  

As shown in the previous section, these G-band bright points tend to have higher magnetic flux density in these magnetic islands.
Furthermore, the minimum magnetic flux density of the G-band bright points becomes higher with increasing distance from the boundary.

The Dopplergram (fig.\ref{plage_dop}) clearly shows an apparent correlation between the magnetic flux density and the suppression of convection.
(The regions enclosed by the contours are the magnetic islands.)
The histograms of the Doppler velocity are shown in figure \ref{dophist1} for the whole region, the G-band bright points and the magnetic islands without the G-band bright points.
The latter two distributions are almost the same. 
These histograms show that suppressed convective velocity of the magnetic islands does not depend on whether a particular magnetic element becomes a G-band bright point or not.

\begin{figure}
\resizebox{9cm}{!}{\includegraphics{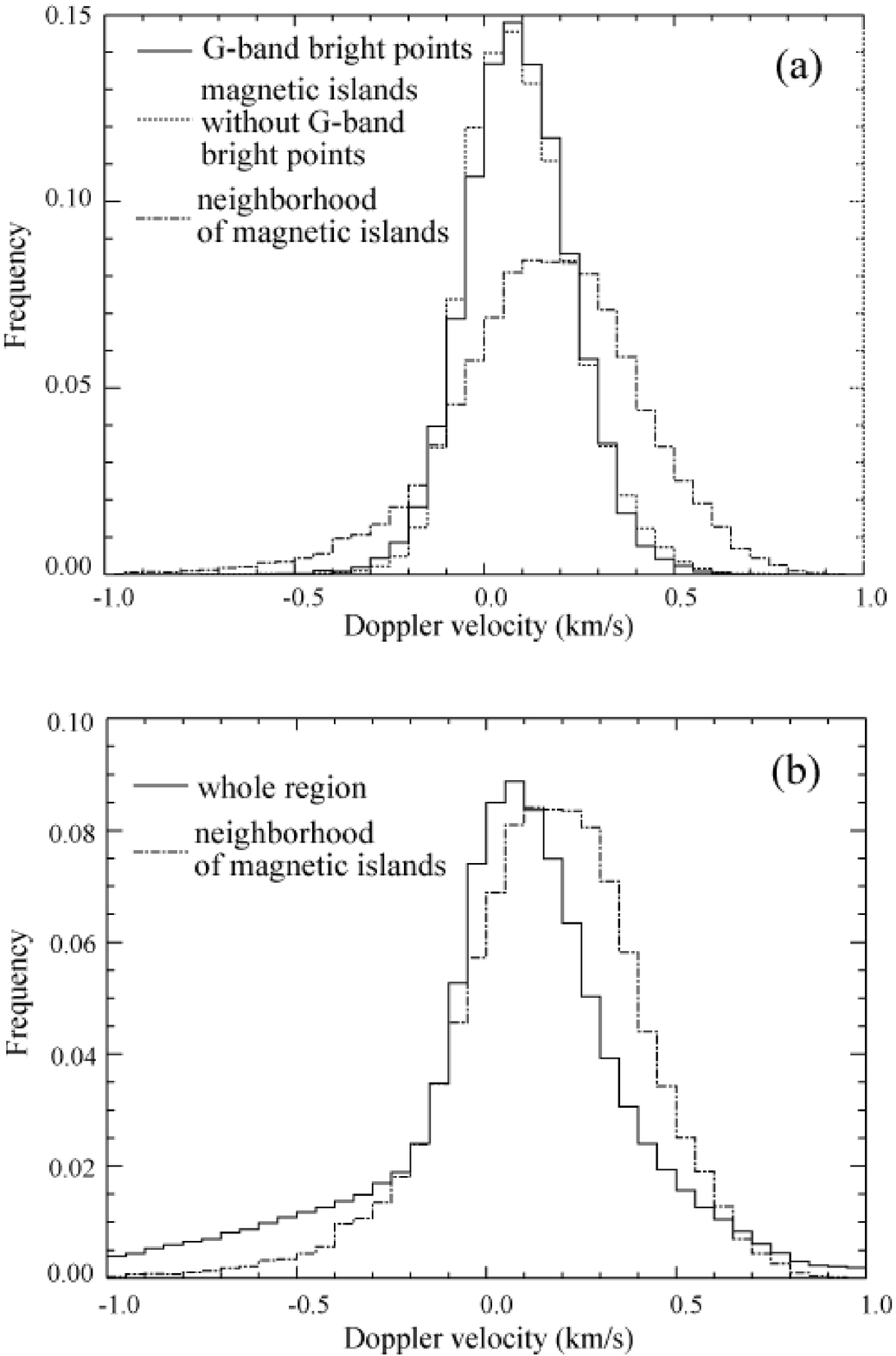}}
  \caption{Histograms of Doppler velocities for different areas. Negative values are upflows (blue-shifted), positive downflows (red-shifted).}
  \label{dophist2}
\end{figure}  

We now examine the environment of the magnetic islands.
By environment, we refer to the surrounding of magnetic islands 1~pixel ($0\farcs04$) to 3~pixels ($0\farcs12$) outward from the boundary of magnetic islands. 
The histograms of Doppler velocity for the environment is shown with those for the G-band bright points, magnetic islands without G-band bright points, and the whole region in figure \ref{dophist2}.
It is clear from these histograms that the environment of the magnetic islands tends to have downflows in comparison with the whole region or with magnetic islands. 
The magnetic islands are surrounded by the thin region (layer), which has predominantly downflows.

\section{Conclusions}
Continuous extended regions with high magnetic flux density are referred to as magnetic islands in this paper.
G-band bright points are a subset of a magnetic island.
The G-band bright points are not recognized in the Dopplergrams: convection is strongly suppressed in any magnetic region, regardless of whether they are associated with the G-band bright points or not.

Our analysis indicates that whether a specific magnetic element is bright in the G-band or not depends on its inward distance from the boundary of the magnetic island as well as on its intrinsic magnetic flux density.
Indeed, $89\%$ of the G-band bright points in magnetic islands are located within $0\farcs41$ from the boundary of the magnetic islands.
This shows that the short distance from the boundary is critically important for the appearance of the G-band bright points. 
Convective downflows close to the boundary would serve as the heat reservoir, and the horizontal radiative energy transport from the reservoir would contribute to the appearance of the G-band bright points.
Interestingly, the inward distance ($\sim 0\farcs12$ $\sim 90$ km) at which the area density of the G-band bright points becomes maximum is comparable to the  horizontal distance for $\Delta\tau$(500nm) $=$ 1 ($\sim$100km).
Moreover, G-band bright points near the boundaries have higher magnetic flux compared with the pixels without G-band bright points.
A stronger magnetic field would decrease the internal plasma pressure, and make $\tau\sim1$ layer deeper in the atmosphere. 
Flux tubes near the boundary are heated by lateral heatflow from the convection outside the magnetic islands.
The heated flux tubes can be visible as the G-band bright points due to the offset of $\tau\sim1$ layer \citep{spruit1976}.

Some pixels, which have very high magnetic flux density compared with that for the G-band bright points near the boundary, are bright in the G-band, even if they are away from the boundary.
In other words, when G-band bright points are located away from the boundaries, higher flux density is needed. 
This is consistent with the above interpretation that strong magnetic field affects $\tau\sim1$ height, and results in G-band bright points.
However, these internal G-band bright points do not appear to have an energy source due to the apparent suppression of vertical convective energy transport.
The magnetic islands may consist of many unresolved elemental magnetic fields, and have vertical and horizontal leakage heat transport.
Another possibility is that the heat leakage comes from a gap (non-magnetized region) underneath the photosphere in the magnetic islands \citep{parker1979, spruit2006}.

This paper addresses the basic question on what makes magnetic foot points bright in the G-band from the observational point of view.
G-band bright points need both the heat transport (horizontal or vertical) path and very strong magnetic field.
G-band bright points have been regarded as good tracers of elemental magnetic fields.
We point out that G-band bright points represent only a subset of magnetic elements, and direct high-resolution observations of the magnetic fields are critical for the understanding of solar magnetic fields.

We also discover that the immediate surrounding of magnetic islands has downflows.
\citet{rouppe2005} and \citet{langangen2007} report that the extended magnetic structures, which they refer to as ribbons, have strong downflows at their edges.
In this paper, we show that the even more extended regions with high magnetic flux density, which are referred to as magnetic islands, have downflows in their immediate surroundings.
These surrounding downflows may be related to the stabilization of the extended strong magnetic region.
Convection would potentially affect not only thin magnetic elemental structures but also the extended magnetic structures.
\begin{acknowledgements}
We thank T. Tarbell, L. Bellot Rubio, and A. Kosovichev for their valuable comments and encouragement. 
The Swedish 1-m Solar Telescope is operated on the island of La Palma by the Institute for Solar Physics of the Royal Swedish Academy of Sciences in the Spanish Observatorio del Roque de los Muchachos of the Instituto de Astrof{\'\i}sica de Canarias.
We thank the RIKEN Super Combined Cluster (RSCC) for the computational resources.\end{acknowledgements}
\bibliographystyle{aa}
\bibliography{6942}

\begin{thebibliography}{28}
\expandafter\ifx\csname natexlab\endcsname\relax\def\natexlab#1{#1}\fi

\bibitem[{{Beckers} \& {Schr{\"o}ter}(1968)}]{beckers1968}
{Beckers}, J.~M. \& {Schr{\"o}ter}, E.~H. 1968, \solphys, 4, 142

\bibitem[{{Berger} {et~al.}(1998){Berger}, {Loefdahl}, {Shine}, \&
  {Title}}]{berger1998}
{Berger}, T.~E., {Loefdahl}, M.~G., {Shine}, R.~S., \& {Title}, A.~M. 1998,
  \apj, 495, 973

\bibitem[{{Berger} {et~al.}(2004){Berger}, {Rouppe van der Voort},
  {L{\"o}fdahl}, {Carlsson}, {Fossum}, {Hansteen}, {Marthinussen}, {Title}, \&
  {Scharmer}}]{berger2004}
{Berger}, T.~E., {Rouppe van der Voort}, L.~H.~M., {L{\"o}fdahl}, M.~G.,
  {et~al.} 2004, \aap, 428, 613

\bibitem[{{Berger} {et~al.}(1995){Berger}, {Schrijver}, {Shine}, {Tarbell},
  {Title}, \& {Scharmer}}]{berger1995}
{Berger}, T.~E., {Schrijver}, C.~J., {Shine}, R.~A., {et~al.} 1995, \apj, 454,
  531

\bibitem[{{Berger} \& {Title}(2001)}]{berger2001}
{Berger}, T.~E. \& {Title}, A.~M. 2001, \apj, 553, 449

\bibitem[{{Bovelet} \& {Wiehr}(2003)}]{bovelet2003}
{Bovelet}, B. \& {Wiehr}, E. 2003, \aap, 412, 249

\bibitem[{{Carlsson} {et~al.}(2004){Carlsson}, {Stein}, {Nordlund}, \&
  {Scharmer}}]{carlsson2004}
{Carlsson}, M., {Stein}, R.~F., {Nordlund}, {\AA}., \& {Scharmer}, G.~B. 2004,
  \apjl, 610, L137

\bibitem[{{Dunn} \& {Zirker}(1973)}]{dunn1973}
{Dunn}, R.~B. \& {Zirker}, J.~B. 1973, \solphys, 33, 281

\bibitem[{{Kano} \& {Tsuneta}(1995)}]{kano1995lp}
{Kano}, R. \& {Tsuneta}, S. 1995, \apj, 454, 934

\bibitem[{{Keller}(1992)}]{keller1992n}
{Keller}, C.~U. 1992, \nat, 359, 307

\bibitem[{{Keller} {et~al.}(2004){Keller}, {Sch{\"u}ssler}, {V{\"o}gler}, \&
  {Zakharov}}]{keller2004}
{Keller}, C.~U., {Sch{\"u}ssler}, M., {V{\"o}gler}, A., \& {Zakharov}, V. 2004,
  \apjl, 607, L59

\bibitem[{{Langangen} {et~al.}(2007){Langangen}, {Carlsson}, {Rouppe van der
  Voort}, \& {Stein}}]{langangen2007}
{Langangen}, {\O}., {Carlsson}, M., {Rouppe van der Voort}, L., \& {Stein},
  R.~F. 2007, \apj, 655, 615

\bibitem[{{Leenaarts} {et~al.}(2006){Leenaarts}, {Rutten}, {Carlsson}, \&
  {Uitenbroek}}]{leenaartsetal2006}
{Leenaarts}, J., {Rutten}, R.~J., {Carlsson}, M., \& {Uitenbroek}, H. 2006,
  \aap, 452, L15

\bibitem[{{Mehltretter}(1974)}]{mehl1974}
{Mehltretter}, J.~P. 1974, \solphys, 38, 43

\bibitem[{{Nisenson} {et~al.}(2003){Nisenson}, {van Ballegooijen}, {de Wijn},
  \& {S{\"u}tterlin}}]{nisenson2003}
{Nisenson}, P., {van Ballegooijen}, A.~A., {de Wijn}, A.~G., \&
  {S{\"u}tterlin}, P. 2003, \apj, 587, 458

\bibitem[{{Parker}(1979)}]{parker1979}
{Parker}, E.~N. 1979, \apj, 234, 333

\bibitem[{{Rouppe van der Voort} {et~al.}(2005){Rouppe van der Voort},
  {Hansteen}, {Carlsson}, {Fossum}, {Marthinussen}, {van Noort}, \&
  {Berger}}]{rouppe2005}
{Rouppe van der Voort}, L.~H.~M., {Hansteen}, V.~H., {Carlsson}, M., {et~al.}
  2005, \aap, 435, 327

\bibitem[{{S{\'a}nchez Almeida} {et~al.}(2001){S{\'a}nchez Almeida}, {Asensio
  Ramos}, {Trujillo Bueno}, \& {Cernicharo}}]{sanchezal2001}
{S{\'a}nchez Almeida}, J., {Asensio Ramos}, A., {Trujillo Bueno}, J., \&
  {Cernicharo}, J. 2001, \apj, 555, 978

\bibitem[{{Scharmer} {et~al.}(2003{\natexlab{a}}){Scharmer}, {Bjelksj{\"o}},
  {Korhonen}, {Lindberg}, \& {Petterson}}]{scharmer2003SST}
{Scharmer}, G.~B., {Bjelksj{\"o}}, K., {Korhonen}, T.~K., {Lindberg}, B., \&
  {Petterson}, B. 2003{\natexlab{a}}, in Proc. SPIE. 4853, Innovative
  Telescopes and Instrumentation for Solar Astrophysics. ed. S.L. Keil \& S.V.
  Avakyan, 341--350

\bibitem[{{Scharmer} {et~al.}(2003{\natexlab{b}}){Scharmer}, {Dettori}, \&
  {L{\"o}fdahl}}]{scharmer2003AO}
{Scharmer}, G.~B., {Dettori}, P.~M., \& {L{\"o}fdahl}, M.~G. an d~{Shand}, M.
  2003{\natexlab{b}}, in Proc. SPIE. 4853, Innovative Telescopes and
  Instrumentation for Solar Astrophysics. ed. S.L. Keil \& S.V. Avakyan,
  370--380

\bibitem[{{Spruit}(1976)}]{spruit1976}
{Spruit}, H.~C. 1976, \solphys, 50, 269

\bibitem[{{Spruit}(1977)}]{spruit1977}
{Spruit}, H.~C. 1977, \solphys, 55, 3

\bibitem[{{Spruit} \& {Scharmer}(2006)}]{spruit2006}
{Spruit}, H.~C. \& {Scharmer}, G.~B. 2006, \aap, 447, 343

\bibitem[{{Steiner} {et~al.}(2001){Steiner}, {Hauschildt}, \&
  {Bruls}}]{steiner2001}
{Steiner}, O., {Hauschildt}, P.~H., \& {Bruls}, J. 2001, \aap, 372, L13

\bibitem[{{Title} \& {Rosenberg}(1981)}]{title1981}
{Title}, A. \& {Rosenberg}, W. 1981, in Solar instrumentation: What's next?,
  ed. R.~B. {Dunn}, 326--+

\bibitem[{{Uitenbroek} \& {Tritschler}(2006)}]{uitenbroektritschler2006}
{Uitenbroek}, H. \& {Tritschler}, A. 2006, \apj, 639, 525

\bibitem[{{van Noort} {et~al.}(2005){van Noort}, {Rouppe van der Voort}, \&
  {L{\"o}fdahl}}]{noort2005}
{van Noort}, M., {Rouppe van der Voort}, L., \& {L{\"o}fdahl}, M.~G. 2005,
  \solphys, 228, 191

\bibitem[{{Wiehr} {et~al.}(2004){Wiehr}, {Bovelet}, \& {Hirzberger}}]{wieh2004}
{Wiehr}, E., {Bovelet}, B., \& {Hirzberger}, J. 2004, \aap, 422, L63

\end{thebibliography}
\end{document}